\documentstyle[12pt]{article}
\begin{document}
\bibliographystyle{unsrt}

\title{Do we have three $S_{11}$ resonances in the second resonance
region?}
\author{Zhenping Li$^1$ and Ron Workman$^2$
\\
$^1$Physics Department, Carnegie-Mellon University
\\
Pittsburgh, PA. 15213-3890
\\
$^2$Department of Physics \\
 Virginia Polytechnic Institute and State
University \\
 Blacksburg, VA. 24061}

\maketitle

\begin{abstract}

We review the status of the $S_{11}$ N$^*$ resonances in light of some
recent theoretical and phenomenological results. Whereas the
quark model predicts two such resonances around 1.6 GeV, there is
considerable evidence for a third  $S_{11}$  resonance
in this energy range.  This suggests that
a $K\Sigma$ or $K\Lambda$ quasi-bound state may indeed exist below the
kaon production threshold. We show that
kaon production experiments, in particular $K^0$  photoproduction
off nucleons, would be very sensitive to the existence of the third
$S_{11}$ resonance.

\end{abstract}


\newpage

There has been considerable progress recently in the investigation of
pion-nucleon scattering and meson-photoproduction off nucleons to
extract the properties of baryon resonances.  The partial wave analysis
of $\pi N$ elastic scattering data to 2.1 GeV has been updated by
the VPI group\cite{VPI}, and a coupled channel analysis of $\pi N\to
\eta N$ and $\eta N\to \eta N$ has also been published\cite{svarc}.
In addition to these renewed analyses, new experimental data for
$\eta$ photoproduction in the threshold region from Bates\cite{dytman},
the Bonn accelerator ELSA\cite{price}, and the Mainz accelerator
MAMI\cite{krusch} have been published.  These data play a unique role
in extracting the properties of the $S_{11}(1535)$ resonance.
In particular, new data from Mainz provide us with more systematic
information on $\eta$ production near threshold, with much better
energy and angular resolution. This enables us
to determine the properties of the $S_{11}(1535)$ resonance more precisely.
On the theoretical side, a new approach based on the chiral quark model
has been developed for meson photoproduction\cite{zpli94,zpli952,zpli95},
and the particular case of $\eta$ photoproduction off nucleons\cite{zpli952}
has been investigated in this new approach. The chiral quark model for
meson photoproduction starts from the low energy QCD
Lagrangian\cite{MANOHAR} so that the meson-quark interaction is
chiral invariant and the low energy theorem\cite{cgln}
for threshold pion photoproduction is automatically recovered.  This
establishes a connection between the reaction mechanism in
photoproduction and the fundamental theory (QCD).  Perhaps
more importantly, it relates the photoproduction data
directly  to the spin-flavor
structure of baryon resonances.  In this paper, we will focus
on properties of the $S_{11}$ resonances,
and highlight possible physical consequences.

One property of the $S_{11}(1535)$ resonance, determined from
$\eta$ photoproduction, is given by the quantity $\xi$,
\begin{equation}\label{1}
\xi=\sqrt{ \frac {M_Nk \chi_{\eta N}}{qM_R\Gamma_T}}A_{\frac 12}
\end{equation}
where $M_N$ ($M_R$) denotes the mass of the nucleon (resonance),
$k$ and $q$ correspond to the momenta of the incoming photon and the
outgoing meson $\eta$, $\chi_{\eta N}$ is the branching ratio of the
resonance to the $\eta N$ channel, and $\Gamma_T$ and $A_{\frac 12}$
are the total width and the helicity amplitude for the resonance.
A study\cite{muko} by the RPI group shows that this quantity
obtained from the experimental data is model independent, and thus
should be calculated in theoretical investigations.
One advantage of the chiral quark model approach is that the
quantity $\xi$ for the $S_{11}(1535)$ resonance can be directly
related to its underlying spin flavor structure, which is expressed by
the analytical form
\begin{eqnarray}\label{2}
\xi=\sqrt{\frac {\alpha_{\eta}\alpha_e\pi(E^f+M_N)}{M_R^3}}\frac
{C_{S_{11}(1535)}k}{6\Gamma_T}\left [\frac
{2\omega_{\eta}}{m_q}-\frac {2{
q}^2}{3\alpha^2}\left (\frac {\omega_{\eta}}{E^f+M_N}+1\right )\right
] \nonumber \\
\left (1+\frac { k}{2m_q}\right )e^{-\frac {{ q}^2+{
k}^2}{6\alpha^2}},
\end{eqnarray}
where $\omega_{\eta}$ and $E^f$ are the energies of the outgoing
$\eta$ meson and the nucleon,
$m_q=0.34$ GeV is the constituent quark mass, and $\alpha^2=0.16$
GeV$^2$ is the parameter in the harmonic oscillator wavefunction.
The coupling of the $S_{11}(1535)$ to $\eta N$ in Eq. \ref{2} is
determined by the $\eta NN$ coupling constant $\alpha_{\eta}$.  This
provides a consistency condition that must be checked in any microscopic
model of baryon
decay amplitudes,  otherwise, the overall agreement with data from
meson photoproduction would be lost. The coefficient $C_{S_{11}(1535)}$
is equal to unity in the naive $SU(6)\otimes O(3)$ quark model.
Thus the quantity $C_{S_{11}(1535)}-1$ measures a deviation
from the underlying  $SU(6)\otimes O(3)$ symmetry.
Both the $S_{11}(1535)$ and $S_{11}(1650)$ resonances
show a strong configuration mixing in more sophiscated models\cite{ik}.

By treating the coupling constant $\alpha_\eta$, the coefficient
$C_{S_{11}(1535)}$ and the total decay width $\Gamma_T$ as free parameters
and fitting them to the experimental data, we find\cite{zpli952}
\begin{eqnarray}\label{3}
\Gamma_T & = & 198 \quad \mbox{MeV} \nonumber \\
C_{S_{11}(1535)} & = & 1.608 \nonumber \\
\alpha_{\eta} & = & 0.435,
\end{eqnarray}
which gives an excellent fit to the recent Mainz\cite{krusch} data.
The total width, $\Gamma_T$,
obtained from the chiral quark model only differs
from a simple Breit-Wigner parameterization by 4 to 5 MeV\cite{krusch}.
The above results give
\begin{equation}\label{31}
\xi = 0.220 \quad \mbox{GeV}^{-1}.
\end{equation}
This value is in good agreement with results of the RPI group, which used an
effective Lagrangian approach  to fit both old data sets and new data from
the Mainz group\cite{muko,muko1}.  An extraction of the helicity
amplitude $A^p_{1/2}$ from the quantity $\xi$ depends on the $\eta N$
branching ratio $\chi_{\eta N}$,  which is not precisely known at present.
One could use, as a guide,
the result from a recent coupled channel analysis by
Batini\'c {\it et al}\cite{svarc}. There the
branching ratios to $\eta N$ and $\pi N$ channels
\begin{equation}\label{32}
\chi_{\eta N} =0.63 \;\; {\rm and} \;\;  \chi_{\pi N} =0.31,
\end{equation}
were found for the $S_{11}(1535)$ resonance, the latter being in
in good agreement with a result from the VPI group\cite{VPI}.
These lead to the helicity amplitude
\begin{equation}\label{4}
A^p_{1/2}=98.9 \quad 10^{-3} \; \mbox{GeV}^{-1/2}.
\end{equation}
However, the total width $\Gamma_T$ for the
resonance $S_{11}(1535)$ varies significantly when
extracted from recent partial
wave analyses\cite{VPI,svarc} and the $\eta$ photoproduction
data\cite{krusch}.  Therefore,
the helicity amplitude $A_{1/2}$  still cannot be extracted
reliably, as it is proportional to $\sqrt{\frac
{\Gamma_T}{\chi_{\eta N}}}$ for a fixed quantity $\xi$.

There are two $S_{11}$ resonances near 1.6 GeV in the quark model
generated by the $SU(6)\otimes O(3)$ basis.
Discrepancies between the theory, in various quark model
calculations, and the properties of the $S_{11}(1535)$ resonance
have been known for some time.  The helicity amplitude $A^p_{1/2}$ from
quark model calculations has\cite{close,simon} remained near
150 $10^{-3}$ GeV$^{-\frac 12}$, while the branching ratio for the
$S_{11}(1535)$ resonance decaying to $\eta N$ is too
small\cite{zpli952,simon}.  The solution within the quark model has been
configuration mixing between the two $S_{11}$ $SU(6)\otimes O(3)$
states.   However, our investigation of $\eta$ photoproduction
indicates that configuration mixing alone may not be enough to
resolve this problem.  The coefficient
$C_{S_{11}(1535)}$, from fits to $\eta$ photoproduction data
in Ref. \cite{zpli952}, is found to be 1.5$\sim$1.6, suggesting
that the naive quark model would predict
\begin{equation}\label{41}
\xi \approx 0.14 \quad \mbox{GeV}^{-1}.
\end{equation}
Notice that the quantity $\xi$ is proportional to the product of the
helicity amplitude $A_{1/2}$ and the meson decay amplitude
\begin{equation}\label{5}
\xi \propto \langle N |H_{\eta}| S_{11}\rangle \langle S_{11}| H_{em}| N
\rangle .
\end{equation}
Configuration mixing effects for the wavefunction of the
$S_{11}$ resonance and the nucleon are unlikely to
increase the quantity $\xi$.
This shows that the branching ratios for the $S_{11}(1535)$ resonance
have not been understood in the
quark model.   For the resonance $S_{11}(1650)$, the naive quark model
predicts
\begin{equation}\label{51}
\xi_{S_{11}(1650)}=0.
\end{equation}
This is consistent with the $\eta$ photoproduction data but for the
wrong reason.  Eq. \ref{51} comes from the Moorhouse selection
rule\cite{moor} which predicts that the electromagnetic
transition between the
nucleon and the $S_{11}(1650)$ resonance vanishes. In fact, while
partial-wave
analyses show that the $\eta N$ branching ratio is very small,
the helicity amplitude for the $S_{11}(1650)$ is substantial.
Both the Kent State\cite{Kent} and Virginia Tech\cite{VPI}
analyses imply that the $S_{11}(1650)$ resonance has
a $\Gamma_{\pi N} / \Gamma$
branching ratio approaching unity, making it the most elastic resonance
apart from the $P_{33}$(1232), a result consistent with the
recent coupled channel analysis of Ref.\cite{svarc}.

Therefore, the enhancement of the $S_{11}(1535)$ resonance and the
suppression of the $S_{11}(1650)$ resonance in the $\eta N$ channel are
certainly key to our understanding of their underlying spin-flavor
structure.  This has motivated a number of different approaches to
the problem.
The investigation by Kaiser {\it at al.}\cite{chiral} has indicated that
a quasi-bound $K\Sigma$ state with properties remarkably similar to the
$S_{11}(1535)$  may be responsible for the large $\eta N$ branching ratio
attributed to the $S_{11}(1535)$ resonance.
In the study of Ref.\cite{chiral}, an
$SU(3)$ effective chiral Lagrangian was applied to the S-wave meson-baryon
interaction and the parameters were determined by low-energy
$\bar {K}N$ experimental data.  The resonance $\Lambda(1405)$\cite{Dalitz}
has been suggested as a possible $K$-nucleon  bound state
whose mass is just below the $KN$ threshold.
It is certainly possible that there is a weakly bounded $K\Lambda$ or
$K\Sigma$ state whose mass lies just below the $K\Lambda$ or $K\Sigma$
threshold.   The problem with this approach is that
data for the electromagnetic transition, in particular the $Q^2$
dependence\cite{stoler} of the helicity amplitude $A^p_{1/2}$, indicates
that the $S_{11}(1535)$ should be a dominantly 3-quark state
at higher $Q^2$, according to the perturbative QCD counting
rule\cite{carlson}.
Thus, if such a quasi-bound $K\Sigma$ state exists,
it should be strongly mixed with the 3-quark configurations.
This requires an additional $S_{11}$ resonance with a mass
near the two known $S_{11}$ resonances predicted by the quark model.

Therefore,  a quasi-bound
$K\Lambda$ or $K\Sigma$ state is possible only if a third
$S_{11}$ resonance exists near the two known $S_{11}$ resonances,
and there is considerable experimental evidence for this
extra $S_{11}$.  The most recent VPI
analysis of $\pi N$ elastic scattering\cite{VPI} claims some evidence
for a third $S_{11}$ resonance with mass and decay width
\begin{equation}\label{6}
M_{S_{11}}=1.712  \quad \Gamma_{S_{11}}=0.184
\end{equation}
in GeV units.  A similar structure was found\cite{hobo}
by H\"ohler in his speed plot of the KA84 solution, although he did not
consider this to be a resonance.  In addition to the analysis of
$\pi N$ elastic scattering data, the coupled channel
analysis of $\pi N\to \eta N$ and $\eta N\to \eta N$\cite{svarc}
results in a solution consistent with the presence of a third $S_{11}$
resonance  with a mass of $1.705$ GeV
and total width $0.27$ GeV. (This state was identified as the
$S_{11}(2090)$.)
In fact, this new $S_{11}$ resonance may already have been seen in a
much older partial wave analysis of the reaction
$\pi N\to K\Sigma$\cite{Deans}, where the $S_{11}$ resonance had a
fitted mass of $1.70 \sim 1.75$ GeV and a
total width $0.210\sim 0.270$ GeV.

More experimental evidence is certainly needed to establish
this resonance.  As the new state is only
slightly above the $S_{11}(1650)$, it would be difficult to determine
which is contributing to a particular reaction. However, if the VPI
and Kent State\cite{VPI,Kent} analyses are correct,
the $S_{11}(1650)$ resonance
should not contribute strongly to reactions having initial and final
states other than $\pi N$, a view supported by the analysis of Batini\'c
{\it et al.}\cite{svarc}.   Furthermore, the partial wave analysis of
$\pi N \to K\Sigma$
by Deans {\it et al.}\cite{Deans} suggests that the coupling of
this resonance
to the $K\Sigma$ final state is quite strong.   Thus kaon production
experiments, such as $\pi N\to K Y$, and $\gamma N\to K Y$, would be
important in establishing its existence.

In addition to a strong coupling to the $K\Sigma$
final state,  the contribution from this resonance would be further
enhanced by threshold effects.  To understand why this is the case,
we show the relation between the masses of the
S-wave resonances, which include $\Lambda(1405)$, $S_{11}(1535)$,
$S_{11}(1650)$ and $S_{11}(1710)$, and the threshold energies of
$\eta$ and kaon production, which include $K N$, $\eta N$, $K\Lambda$
and $K\Sigma$ channels in Fig. 1.  The S-wave resonances are clearly
sandwiched between the threshold energies of $\eta$ and $K$ productions.
Our investigations of meson photoproduction show\cite{zpli952,zpli95}
that there are two major factors determining the threshold behaviour,
the leading Born terms which are dominated by the contact (seagull)
term, and the S-wave resonances whose masses are near the threshold
energies shown in Fig. 1.  The contact term is proportional to the
charge of the photoproduced meson and the form factors, generated by
the spatial wavefunctions, predict a forward peaking behavior.
This is confirmed by data\cite{sighai} for the $\gamma p\to
K^+\Lambda$ reaction.  On the other hand, the leading contact
term does not contribute to the photoproduction of neutral mesons,
thus the role of S-wave resonances near the threshold region is
enhanced.  This is one of the major reasons that the
$S_{11}(1535)$ resonance dominates the threshold region in
$\eta$ photoproduction.
The calculation of $\gamma p\to \eta p$\cite{zpli952} shows that the
$S_{11}(1535)$ resonance still dominates the threshold region even with
a smaller $\xi$ consistent with the quark model prediction.
Note that the resonances $S_{11}(1535)$ and $S_{11}(1710)$ are just above
the $\eta N$ and $K\Lambda$ or $K\Sigma$ threshold energies respectively.
A similar behavior should be found in $K^0$ production, such
as $\gamma n\to K^0\Lambda$ and $\gamma p\to K^0 \Sigma^+$,
in which the resonance $S_{11}(1710)$ would play the same role in
$K^0$ photoproduction as the $S_{11}(1535)$ resonance in $\eta$
photoproduction.  Therefore, we expect that the
$S_{11}(1710)$ resonance will be enhanced in threshold $K^0$ production
if such a state exists. More precise data for $\pi N\to
K Y$ would also be needed for a systematic study of its
properties.

In summary, we have shown that there is considerable evidence
suggesting the existence of a third $S_{11}$ resonance with a mass near
1.7 GeV.  This might be the key to our understanding of the enhancement
of the $S_{11}(1535)$ and suppression of the
$S_{11}(1650)$  in the $\eta N$ channel.  We suggest that a third
$S_{11}$ resonance would support
the existence of a quasi-bound $K\Sigma$ or $K\Lambda$
state, and kaon production experiments, in particular
$\gamma p\to K^0\Sigma^+$ and $\gamma n\to K^0\Lambda$, $K^0\Sigma^0$,
would be very sensitive to this resonance.
The existence of a third $S_{11}$ resonance would certainly provide
challenges to the theory, suggesting that the quark model must incorporate
chiral dynamics in order to provide a consistent treatment of
the $S_{11}$ resonances.   Future experiments
at CEBAF\cite{shoemaker} and ELSA will provide us more
information in this regard.

This work was supported in part by the U.S. Department of
Energy Grant DE-FG05-88ER40454, and the U.S.
National Science Foundation grant PHY-9023586.

\newpage

\subsection*{Figure Caption}
\begin{enumerate}
\item Location of S-wave resonances and the threshold energies
for $K N$, $\eta N$, $K\Lambda$ and $K\Sigma$ production.
\end{enumerate}

\end{document}